\documentstyle[aps,preprint,prb,epsf]{revtex}
\begin{document}                        
\title{A Fault-Tolerant Superconducting Associative Memory}
\author{P. Chandra$^1$ and L.B. Ioffe$^{2,3}$}
\address{$^1$NEC Research Institute, 4 Independence Way, Princeton NJ
08540}
\address{$^2$Landau Institute for Theoretical Physics, Moscow, RUSSIA}
\address{$^3$Department of Physics, Rutgers University, Piscataway, NJ
08855}
\maketitle
\vskip 0.5 in

%
%
 
\textbf{
The demand for high-density data storage with ultrafast accessibility
motivates the search for new memory implementations.  Ideally such
storage devices should be robust to input error and to unreliability
of individual elements; furthermore information should be addressed
by its content rather than by its location.  Here we present a
concept for an associative memory whose key component is a
superconducting array with natural multiconnectivity.  Its intrinsic 
redundancy is crucial for the content-addressability of the
resulting storage device and also leads to parallel image retrieval.  
Because patterns are stored nonlocally both physically and
logically in the proposed device,
information access and retrieval are fault-tolerant.  This superconducting
memory should exhibit picosecond single-bit acquisition times with
negligible energy dissipation during switching and multiple
non-destructive read-outs of the stored data.}

The key component of our proposed associative memory is a
superconducting array with multiple interconnections (Figure 1),
where each bit is represented by a wire and thus is physically
delocalized.
More specifically this network consists of a stack of two
perpendicular sets of $N$ parallel wires separated by a thin oxide
layer.\cite{Vinokur87,Chandra95}  
At low temperatures a superconductor-insulator-superconductor
layered structure, known as a Josephson 
junction,\cite{Josephson64,Anderson67} exists at each node
of this array; logically
each pattern in our proposed memory
is stored nonlocally in these $N^2$ interconnections.  
We note that in this network each horizontal/vertical
wire is coupled to each vertical/horizontal one by a Josephson
junction so that in the thermodynamic limit ($N \rightarrow \infty$
for fixed area) the number of neighbors diverges. 

The important energy-scales of this long-range array are those
associated with the superconducting wires and with the Josephson
junctions.  Each superconducting wire is characterized
by a macroscopic phase
which is constant in equilibrium;
here we assume that phase slips in each wire are energetically unfavorable.
Application of a magnetic field results in the rotation of
this phase, where the rotation rate is determined by the
amplitude of the applied field.\cite{Tinkham80}  The interaction energy of
a Josephson junction is minimized when the phase difference
across its insulating layer is zero. In the absence of
a field this condition is satisfied at each junction of
the array.  However application of a field transverse to the
network results in an overconstrained system since the
$2N$ phases and the $N^2$ Josephson junctions have competing
energetic requirements. The identification of the ground-state
in such a system is a hard combinatorial optimization 
problem,\cite{Mezard87,Monassen99}
as the number of metastable states scales exponentially\cite{Chandra98} 
with the number of wires, $N$. 

Because of its high-connectivity, the long-range Josephson array is
accessible to analytic treatment; furthermore it can be fabricated and
studied in the laboratory.\cite{Chandra97,Shea97}
A detailed theoretical characterization
of this network has been performed.\cite{Chandra95}  
At low temperatures the system
has an extensive number of states, ${\cal N}_{states} \sim e^{cN}$
where $c \sim O(1)$,
separated by free energy barriers that scale with the number of
wires, $N$.  Its specific low-temperature configuration is determined
by sample history, a feature also observed in glassy materials.
Experiments have confirmed predicted static properties of this
multi-connected array, though detailed dynamical investigations in
the laboratory remain to be performed.\cite{Shea97}  

The proposed superconducting network (Figure 1) has long-range
temporal correlations (memory) and an extensive number of
metastable states, and thus it is natural to explore its possible
use for information storage.  Indeed high-connectivity and nonlinear
elements (e.g. Josephson junctions) are key features required
for the construction of associative memories.\cite{Hertz91}  
Here one would
like to store $p$ patterns in such a way that if the memory is exposed
to another one ($\xi_i$)
with a significant
($\ge\frac{1}{\sqrt{N}}$) overlap
with
a stored image ($\tilde{\xi}_i$), 
$q = \frac{1}{N} \sum_i^N \xi_i\tilde{\xi}_i$,
then it produces $\tilde{\xi}$.  A simple model for such a memory is 
based on an array
of McCulloch-Pitts neurons (Figure 2).  The patterns are stored in the
couplings, $J_{ij}$.  Each
nonlinear element has multiple inputs, and the output is a nonlinear
function of the weighted sum of the inputs.  The McCulloch-Pitts
network, with inputs $n_i \in \{0,1\}$, can be reformulated as a spin
model where $\xi_i = 2n_i - 1$; then the output is
\begin{equation}
\xi_i = {\rm sgn} \left( \sum_j J_{ij} \xi_j \right)
\label{xi}
\end{equation}
where 
\begin{equation}
{\rm sgn}(x) = \left\{
\begin{array}{cc}
+1 \quad x\ge 0\\
-1 \quad x\le 0 
\end{array}
\right. 
\label{sgn}
\end{equation}
and the couplings $J_{ij}$ can have arbitrary sign.  Clearly the
output is robust to errors in the input due to the multiple
connections
present. 

In order to ensure that the McCulloch-Pitts array is
content-addressable,
the couplings must be chosen so that the stored images correspond to
stable configurations of the network.  Hopfield has proposed
an algorithm\cite{Hopfield82} where the desired patterns are local minima of an
energy function, e.g. $H = -\frac{1}{2} \sum_{ij} J_{ij}S_i S_j$
where $S_i \in \{-1,+1\}$.
The couplings are chosen so that the energy is minimized for maximal
overlap of $S_i$, the array configuration, and the desired output,
$\xi_i$.
For one pattern, this condition is satisfied for
$J_{ij} = \frac{1}{N} \xi_i \xi_j$ where $N$ is the number of elements
in the array; then $H = -\frac{1}{2N} \left(\sum_i S_i \xi_i
\right)^2$.
We note that with this choice of weights $J_{ij}$ the output is
\begin{equation}
\xi_i = {\rm sgn} \left( \sum_{j=1}^{N} J_{ij} \xi_j \right) 
\label{pattern}
\end{equation}
for all $i$ since $\xi_j^2 = 1$ where $\xi_j$ and $\xi_i$ are
the desired inputs and outputs respectively.  We note that,
according to (\ref{pattern}) with the discussed choice of $J_{ij}$,
a real input $S_j \approx \xi_j$ 
yields the desired output if it has errors in less than
half its bits.
In the 
Hopfield model, the couplings associated with several stored
images are simple superpositions
of the one-pattern case such that
\begin{equation}
J_{ij} = \frac{1}{N} \sum_{\mu = 1}^{p} \xi_i^\mu \xi_j^\mu
\label{hopfield}
\end{equation}
where $\mu$ labels each pattern.  The total storage capacity of
the network, $p_{max}$, is dependent on the acceptable error
rate; in general $p_{max} = \alpha N$ where $\alpha \le 0.138$
if the probability of an erroneous bit in each pattern
is $P_{error} < 0.01$.  Here we discuss the Hopfield
algorithm because of its simplicity, but we note that other
more efficient algorithms\cite{Hertz91} can also be implemented in this
network.

The long-range Josephson array (Fig. 1) can be adapted to become
a superconducting analogue of a McCulloch-Pitts network.  It
is described by the Hamiltonian
\begin{equation}
{\cal H} = {\rm Re} \sum_{j \bar{j}} S_j^* J_{j \bar{j}} S_{\bar{j}}
\label{H}
\end{equation}
with $1 \le (j, \bar{j}) \le N$ where $j$ and $\bar{j}$ are the indices
of the horizontal and vertical wires respectively; $S_j$ are effective
complex spins with unit amplitude, 
$S_j = e^{i\phi_j}$ where $\phi_j$ is the phase of
the $j$-th superconducting wire.  The couplings are site-dependent
and are related to the enclosed flux, $\Phi_{j\bar{j}}$,
in a given area whose edges are defined by wires $j$ and $\bar{j}$
such that
\begin{equation}
J_{j\bar{j}} = \frac{J}{\sqrt{N}} \exp {\frac{2\pi i
\Phi_{j\bar{j}}}{\Phi_0}}.
\label{coupling}
\end{equation}
where $\Phi_0$ is the flux quantum.
For a uniform magnetic field $H$, $\Phi_{j\bar{j}} = H(\bar{j}jl^2)$
where $l$ is the interwire spacing.
We emphasize that the sign of $J_{j\bar{j}}$
in (\ref{coupling}) can be both positive and negative
depending on the value of $\frac{\Phi_{\j\bar{j}}}{\Phi_0}$.  
In complete analogy with our previous discussion of the
McCulloch-Pitts
network, patterns are stored in this superconducting associative
memory
by appropriate choice of the weights, $J_{j\bar{j}}$.  Because the 
$J_{j\bar{j}}$
are functions of the enclosed fluxes, $\Phi_{j\bar{j}}$, they can
be set to their desired values by appropriately tuning the local applied field
$H_{j\bar{j}}$.
In practice this writing procedure could be accomplished by
an array of superconducting quantum interference
devices (SQUIDs) superimposed on the multi-connected Josephson
network.\cite{Chandra97b}

The stored patterns are encoded in the Josephson couplings of the
long-range array and correspond to stable configurations of the
$2N$ superconducting phases.  A ``fingerprint''
of each image can then be determined using voltage pulses and
the Josephson phase-voltage relation 
$\Delta \phi = \frac{2\pi}{\Phi_0} \int V dt$ where $\Delta \phi$
is the phase difference across the relevant 
junction.\cite{Josephson64,Anderson67}  
More specifically
a set of voltage pulses can be applied to a small subset ($ >
\sqrt{N}$)
of the horizontal
wires, a ``key'',  thereby altering the phase differences at the
associated
nodes. 
The phases
of the vertical wires must readjust in order for the system
to settle into a stable configuration, a process which
results in the absence/presence of a voltage pulse.
The set of key input and $N$ output voltage pulses therefore
constitutes a signature of each stored image.
Single-flux quantum (SFQ) voltage pulses,where 
$\int V dt = \Phi_0$, may be used for direct
analogy with the McCulloch-Pitts array
where inputs $n_{\bar{j}} \in \{0,1\}$ now refer to the absence/presence of
a SFQ pulse.  Again it is convenient to describe the  network
in terms of the variables $\xi_i = 2n_i - 1$. Following the
Hopfield algorithm, the local coupling associated with
one pattern is
$J_{j\bar{j}} = \frac{1}{N} \xi_j \xi_{\bar{j}}$
in accordance with (\ref{coupling})
so the the desired output $\xi_j = J_{j\bar{j}} S_{\bar{j}}$ is robust
for $S_{\bar{j}} \approx \xi_{\bar{j}}$.  
The weights coding
many stored patterns are superpositions, (\ref{hopfield}),
of the
one-pattern cases.
From a practical
standpoint, these couplings are ``written'' by local (uniform)
fields applied to individual plaquettes of area $l^2$; for a cell
with its lower left-hand corner defined by the Cartesian coordinates
$(j,\bar{j})$, the plaquette flux,
$\Phi_{plaquette}^{j\bar{j}}$,
is related to the weights by the expression
\begin{equation}
\Phi_{plaquette}^{j\bar{j}} = \frac{\Phi_0}{2} 
\Theta \left\{-J_{j\bar{j}}J_{j+1\bar{j}}J_{j+1\bar{j}+1}J_{j\bar{j}+1}
\right\}
\label{flux}
\end{equation}
where $\Theta(x) = 1$ if $x \ge 0$ and $\Theta(x) = 0$ otherwise.  
Again we note that other algorithms can be used to determine the
couplings;
here we use 
the Hopfield model as an illustrative example.

In the proposed superconducting memory, each stored image is coded
by a distinct set of superconducting phases associated with weak
supercurrents and negligible induced fields. 
In conventional superconducting memories/logic,
digital information is stored locally in
trapped magnetic fluxes
that are switched between SFQ states, $\Phi \in \{0,\Phi_0\}$, 
by Josephson junctions;\cite{Fulton73,Likharev93}
therefore the associated supercurrents should be large
and can lead to unwanted crosstalk between adjacent elements.
However in order to maintain their advantage in speed
compared to other memory technologies, Josephson junction devices
must use SFQ for both information storage and retrieval.  
This condition is satisfied by the design of our
proposed memory where the Josephson junctions 
switch fluxoids while the applied magnetic fluxes
remain fixed; it is not the local fields but the supercurrents
that store the information.

The fault-tolerance of the long-range Josephson network discussed here is due
to the nonlocal nature of its data storage both at the physical
and the logical levels. In conventional planar
superconducting arrays there are $O(N^2)$ individual
{\sl short}
superconducting wires and the fluxoids are
spatially confined to areas $A \sim l^2$ where $l$ is the internode
spacing. By
contrast, in the multi-connected network the phases reside on the $2N$
wires of length $Nl$; thus the fluxoids here  are extended to the entire
system.  Data is coded nonlocally in configurations of these superconducting
phases, similar to the situation in an 
optical holographic storage device.\cite{VanHeerden63}  There
the stored patterns are independent of the input and an analogous
superconducting holographic memory can be constructed.  For example,
let us consider the input wavefunction
\begin{equation}
\Psi_{\bar{j}}^p = \exp 2\pi i \phi_{\bar{j}}^p = \exp \frac{2\pi i \bar{j}
p}{N}
\label{hinput}
\end{equation}
where $\bar{j}$ and $p$ are indices labelling the horozontal wires
and the stored patterns respectively. Then the input voltage pulses
would be $\int V dt = \left[ \frac{\bar{j}p}{N}\right] \Phi_0$ where
$[\quad]$ refers to the fractional part.  Using the Hopfield algorithm,
we have $J_{j\bar{j}} = \frac{1}{N} \sum_p \xi_{j}^p \Psi_{\bar{j}}^p$
which yields the desired output $\xi_j^p = \sum_j J_{j\bar{j}}^*
\Psi_{\bar{j}}^p$.  We note that any orthogonal basis for the inputs
will work; therefore this long-range Josephson array can be used
as a key component of both an associative and a holographic memory.
  
Practically, the proposed superconductor memory cell consists of
the superimposed SQUID and long-range networks for writing
and reading respectively, and a phase reset 
circuit. Each data
retrieval event in the READ array must be followed by a reset operation
since the output signals correspond to phase differences with respect
to a reference state.  This reset circuit can be constructed from
a series of double-junction SQUID loops connected to each horizontal
wire of the multi-connected array with a variable coupling to ground;
if finite this coupling locks the relevant phase into a reference
state, whereas if zero (e.g. in the presence of a control line current)
the next data retrieval process can be performed.\cite{Chandra97b}  
This memory cell can then
be embedded in an enviroment with known input/output SFQ circuitry
that includes DC/SFQ voltage pulse converters and SFQ 
transmission/amplification lines.\cite{Likharev93} 
The network parameters, the charging ($E_c$) and
the coupling ($E_J$) energy scales, should be chosen to
optimize performance, particularly to maximize access rate,
$\omega \sim {\rm min}(\Delta, \sqrt{E_cE_J})$ where
$\Delta$ refers to the superconducting gap.  An additional
constraint on $E_C$ ($E_C \leq 0.01 E_J$) results from the condition
that phase fluctuations remain weak and do not result in
errors.  Given these constraints, the minimal dissipation
per bit ($\sim E_J$) is roughly $10 \Delta$ which
is $\sim 10^{-22}$ joule for aluminum in contrast to 
its value of $\sim 10^{-15}$ joule for
conventional semiconducting electronics.\cite{Keyes88}

In summary, we have proposed an associative memory device that is
a superconducting analogue of a McCulloch-Pitts network.  It is
content-addressable with a single-bit data accessible time, $\tau_A$,
that is determined by the superconducting gap and the charging
and coupling energy-scales in the network.
Moreover because
this memory is
intrinsically parallel due to its crossbar design, an image of $N$
bits can be retrieved in a time (per bit) 
$\tau_{DT} \sim \frac{\tau_A}{N}$; by contrast
$\tau_{DT} = \tau_A$ in a conventional local memory. 
For example,
an array of $N=1000$ wires with $l=0.5\mu$ has a capacity of
$C = 0.1N^2 = 10^5$ bits; a set of such arrays on a typical
1cm$^2$ chip would then have a capacity of 1 Gigabyte 
with an image access time (per bit) of
$\tau_{DT} = 10^{-15}$ seconds.
The fault-tolerance of this
superconducting memory enhances its appeal as a 
candidate for ultrafast high-density information storage without
conventional problems of power dissipation and subsequent heat removal.  As
a point of comparison, we remark that this proposed device is faster but has
lower absolute capacity than the best optical holographic
memories;\cite{VanHeerden63} this is
because the latter are intrinsically three-dimensional.  We end by noting
that we can tune the long-range array such that its stored images are
maximally distant
from each other in phase space.  In this case the matrix elements
associated with external noise will be negligible, and these
patterns will have long decoherence times.  Such orthogonal
configurations
could be promising as basis states for quantum computation.

\begin{figure}
\centerline{\epsfxsize=10cm \epsfbox{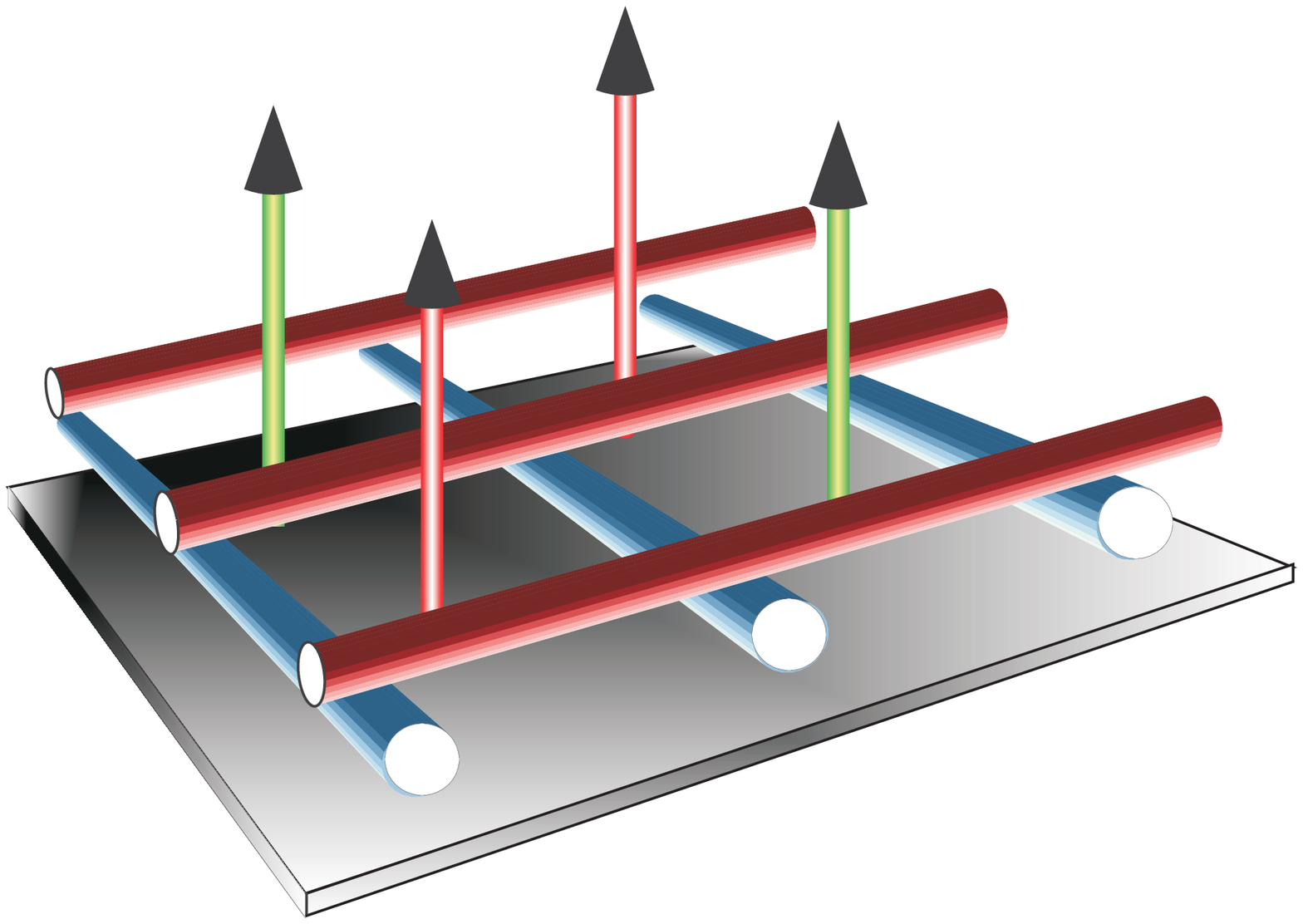}} 
{\footnotesize 
\textbf{Figure 1.}
A schematic of the long-range superconducting array discussed
in the text; the horizontal and vertical wires are coupled
by Josephson junctions at each node and the vertical
arrows refer to the local fields applied to individual
plaquettes that determine the weights. 
}
\end{figure}
\vfill\eject

\begin{figure}
\centerline{\epsfxsize=10cm \epsfbox{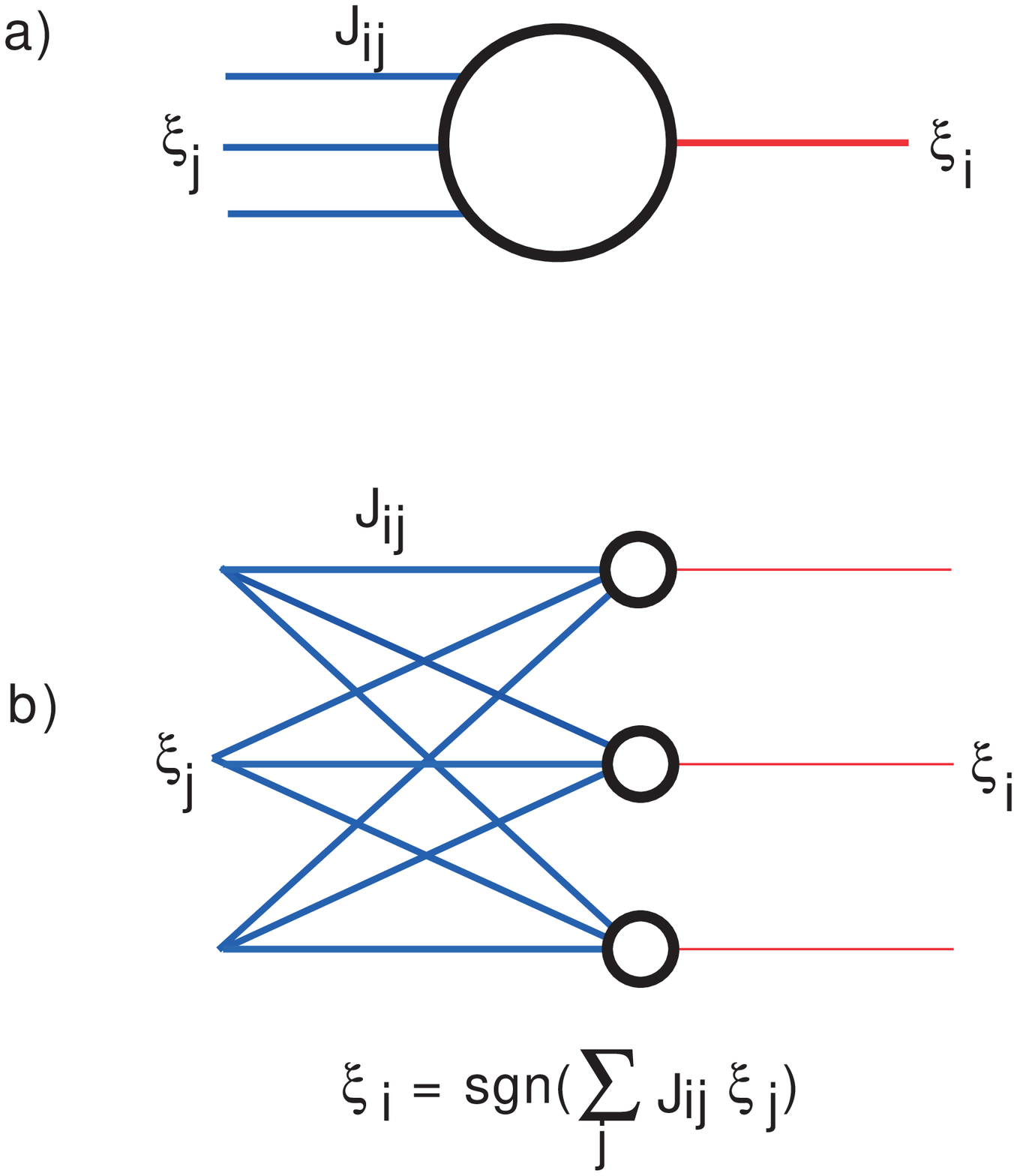}} 
{\footnotesize 
\textbf{Figure 2.}
Schematic diagrams of (a) a McCulloch-Pitts
neuron (b) a McCulloch-Pitts network
where $\xi_i = \pm 1$ and the $J_{ij}$ can have
arbitrary sign. The high-connectivity and the nonlinear
elements in the McCulloch-Pitts array are crucial for
its content-addressability and fault-tolerance.}
\end{figure}

\end{document}